\documentstyle[preprint,aps,epsfig]{revtex}
\begin{document}
\draft
\title{The Mixed-Isospin Vector Current Correlator in Chiral
\\ Perturbation Theory and QCD Sum Rules \\
\begin{flushright}
ADP-95-20/T179 \\
hep-ph/9504237
\end{flushright}}
\author{Kim Maltman\cite{byline}}
\address{Department of Mathematics and Statistics, York University,
4700 Keele St., \\ North York, Ontario, Canada M3J 1P3}
\date{\today}
\maketitle
\begin{abstract}
The mixed-isospin vector current correlator, $\langle 0\vert T(V^\rho_\mu
V^\omega_\nu )\vert 0\rangle$ is evaluated using both QCD sum rules and
Chiral Perturbation Theory (ChPT) to one-loop order.
The sum rule
treatment is a modification of previous analyses necessitated by the
observation that those analyses produce forms of the correlator that
fail to be dominated, near $q^2=0$, by the most nearby singularities.
Inclusion of contributions associated with the $\phi$ meson rectify this
problem. The resulting sum rule fit provides evidence for a
significant direct $\omega\rightarrow\pi\pi$ coupling contribution in
$e^+e^-\rightarrow\pi^+\pi^-$.  It is also pointed out that results for the
$q^2$-dependence of the correlator cannot be used to provide information about
the (off-shell) $q^2$-dependence of the off-diagonal element of the vector
meson propagator unless a very specific choice of interpolating fields for
the vector mesons is made.  The results for the value of the correlator near
$q^2=0$ in ChPT are shown to be more than an order of magnitude smaller
than those extracted from the sum rule analysis and the reasons why this
suggests slow convergence of the chiral series for the correlator given.
\end{abstract}

\section{Introduction}\label{sec:I}

Non-electromagnetic isosopin breaking is well-established in many
strongly interacting systems (e.g., splittings in the hadron spectrum,
binding energy differences in mirror nuclei, asymmetries in polarized
np scattering, binding energies and level splittings of light $\Lambda$
hypernuclei\onlinecite{ref1}).  In few-body systems, an important
source of this breaking has been thought to be the mixing of
isoscalar and isovector mesons appearing in meson exchange
diagrams.  In particular, the bulk of the non-Coulombic contributions
to the charge symmetry breaking $nn$-$pp$
scattering length difference and to the A=3
binding energy difference, and of the np asymmetry at $183$ MeV,
can be explained\onlinecite{ref2,ref3} using the value of $\rho -\omega$
mixing extracted from an analysis of $e^+e^-\rightarrow\pi^+\pi^-$ in
the $\rho -\omega$ interference region\onlinecite{ref4,ref5} .  The
plausibility of this explanation (which employs the observed
mixing, measured at $q^2=m_\omega^2$, unchanged in the
spacelike region $q^2<0$) has, however, recently been called
into question by Goldman, Henderson and Thomas\onlinecite{ref6} who
pointed out that, in the context of a particular model, the
relevant $\rho -\omega$ mixing matrix element has significant
$q^2$-dependence.  Subsequently, various authors, employing various
computational and/or model framewords, have showed that the presence
of such $q^2$-dependence appears to be a common feature of
isospin-breaking in both meson-propagator- and current-correlator
matrix elements
\onlinecite{ref7,ref8,ref9,ref10,ref11,ref12,ref13,ref14,ref15,ref16}~.

In the present paper we will concentrate on the isospin-breaking vector
current correlator
\begin{equation}
\Pi_{\mu\nu}(q^2)= i\int d^4x\ e^{iq.x}
\langle 0\vert T(V^\rho_\mu (x)\, V^\omega_\nu (0))\vert 0\rangle\  ,
\eqnum{1.1}\label{oneptone}
\end{equation}
where
\begin{equation} V^\rho_\mu =(\bar{u} \gamma_\mu u -\bar{d}\gamma_\mu
d)/2\ ,\quad V^\omega_\mu = (\bar{u}\gamma_\mu u+\bar{d}\gamma_\mu d)/6\ .
\eqnum{1.2}\label{onepttwo}\end{equation}

This correlator was first analyzed using QCD sum rules in
Ref.~\onlinecite{ref17}~, and the analysis updated by the
authors of Ref.~\onlinecite{ref12} who, in particular, stressed
the $q^2$-dependence of the correlator implicit in the results
of this analysis.  As will be shown below, a worrisome feature of
the resulting fit is that the phenomenological representation of the
correlator near $q^2=0$ is not dominated by the most nearby singularities,
suggesting that some ingredient may be missing from the form chosen
for this representation.  This missing ingredient is identified
below and it is shown that a reanalysis of the correlator, which includes
it, rectifies the problem.  The resulting correlator still displays
a very strong $q^2$-dependence, and, in addition, provides evidence
for the presence of significant direct $\omega\rightarrow\pi\pi$
coupling in $e^+e^-\rightarrow\pi^+\pi^-$.
The behavior of the resulting correlator near $q^2=0$
is then compared with that
obtained from ChPT to one-loop.  The latter is found to be more
than an order of magnitude smaller than the former, the reason
why this suggests the likelihood of a slow convergence of the chiral series
for the correlator explained.

The paper is organized as follows.  In Section II,
those features of the behavior of quantum field theories under field
redefinitions relevant to attempts in the literature to relate
meson propagators and current correlators are discussed,
and it is explained why the
freedom of field redefinition implies that (1) one cannot obtain off-shell
information about the off-diagonal element of the vector meson
propagator from the off-diagonal element of the vector current
correlator without making specific choices for the vector
meson interpolating fields, and (2) if one writes the off-diagonal
element of the vector meson propagator as
\begin{equation}
\Delta_{\mu\nu}^{\rho\omega}(q^2)\equiv -(g_{\mu\nu}-q_{\mu}q_\nu /q^2)
{\theta^{\rho\omega}(q^2)\over (q^2-m^2_\rho )(q^2 -m^2_\omega )}
\eqnum{1.3}\label{oneptthree}\end{equation}
$\theta^{\rho\omega}(q^2)$ cannot, in general, be $q^2$-independent.
In Section III we return to the QCD sum rule analysis of the vector
current correlator, first explaining why certain features of the
existing analyses suggest the need for a modified analysis,
and then performing this analysis.  The results both correct the
apparently unphysical features of the previous analyses and
provide evidence for non-negligible direct $\omega\rightarrow\pi^+\pi^-$
contributions to $e^+e^-\rightarrow\pi^+\pi^-$ in the $\rho -\omega$
interference region.  In Section IV, the correlator is computed
to one-loop in ChPT, and the results compared to those obtained
from the sum rule analysis.  Implications for the discrepancy
between the results of the two approaches are discussed there.
Finally, in Section V, a brief summary of the main results of the paper
is given.

\section{Consequences of the Freedom of Field Redefinition}\label{sec:II}

Let us begin by clarifying the relation (or lack thereof)
between the vector-meson-propagator and vector-current-correlator matrices.
The
former is an, in general, off-shell Green function, which we may
think of as being associated with some low-energy effective
Lagrangian, ${\cal L}_{eff}$, in which the vector meson degrees
of freedom have been made explicit.  As is
well-known\onlinecite{ref18,ref19,ref20}~,
the form of such a Lagrangian is not unique:
if $\phi$ and $\chi$ are two possible field choices describing
a given particle, related by $\phi =\chi F(\chi )$, with $F(0)=1$,
then ${\cal L}_{eff}[\phi ]$ and ${\cal L}^\prime_{eff}[\chi ]\equiv
{\cal L}_{eff}[\chi F(\chi )]$ produce exactly the same
experimental observables\onlinecite{ref18}~.  However, while the
S-matrix elements of the two theories are identical, this is
not true of the general off-shell Green functions.  One is free to
make field redefinitions of the form above (as is done, e.g., in order
to obtain the canonical form of the effective Lagrangian for
ChPT\onlinecite{ref19,ref20,ref21}~) without changing the
physical consequences of the theory; the Green functions, however,
are not in general invariant under such field redefinitions.  Useful
pedagogical illustrations of this general principle, for pion Compton
scattering and the
linear $\sigma$-model, are given in Ref.~\onlinecite{ref23} and
Chapter IV of Ref.~\onlinecite{ref22}~,
respectively.  In the case of interest to us, what this means is that,
when we make a redefinition of the $\rho$, $\omega$ fields in
${\cal L}_{eff}[\rho ,\omega ]$, we generate a new effective Lagrangian,
${\cal L}^\prime_{eff}[\rho^\prime ,\omega^\prime ]$, the Green
functions of which are, in general, different from those of ${\cal L}_{eff}$
(though when we piece such Green functions together to form
S-matrix elements, these differences produce no net effect).
The off-shell behavior of the vector meson propagator is thus
dependent on the particular choice of fields used to
represent the vector mesons (the choice of ``interpolating field'').
It is not a physical observable.  In contrast, the vector
current correlators $\Pi_{\mu\nu}^{ab}(q^2 )= i\int d^4x\ e^{iq.x}
\langle 0\vert T(V^a_\mu (x)\, V^b_\nu (0))\vert 0\rangle$ are, in fact,
physical objects, independent of interpolating field choice.
The spectral functions for
$\Pi^{33}_{\mu\nu}$ and $\Pi^{88}_{\mu\nu}$ are, for example,
accessible from a combination of $\tau^-\rightarrow\nu_\tau\pi^-\pi^0$
and $e^+e^-\rightarrow\pi\pi \, ,\, \pi\pi\pi\pi$ data, and that for
$\Pi^{38}_{\mu\nu}$ could in principle be obtained from a careful
analysis of the deviation of the ratio of the differential decay
rates for $\tau^-\rightarrow\nu_\tau\pi^-\pi^0$ and
$e^+e^-\rightarrow\pi^+\pi^-$ from that predicted by isospin
symmetry.  As such there can be no general (i.e. valid for all
choices of interpolating field) relation between the correlator
and propagator matrices.  This point is the source of some
confusion in Ref.~\onlinecite{ref12} where an attempt is made
to obtain the off-shell propagator based on an analysis of the
correlator.

Before proceeding to the reanalysis of the correlator, let us
be more precise about the problems with the interpretation of the
results of Ref.~\onlinecite{ref12}, in the light of the above
comments.  The authors begin by writing a general form for the
spectral function of the correlator:
\begin{equation}
{\rm Im}\, \Pi_{\mu\nu} (q^2)=A_0\, {\rm Im}\, \Pi^{\rho\omega}_{\mu\nu}(q^2)
+A_1\, {\rm Im}\, \Pi^{\rho^\prime\omega^\prime}_{\mu\nu}+\cdots
\eqnum{2.1}\label{twoptone}\end{equation}
where the superscripts on the RHS should, for the moment, be
taken only as labelling the region of spectral strength, and
where $+\cdots$ refers to all other contributions (we return to
this below).  Eqn.~(2.1) is, of course, completely
general.  The authors of Ref.~\onlinecite{ref12}, however, then
identify $A_0$ with $m^2_{\rho}m^2_\omega /g_\rho g_\omega$,
where $g_{\rho ,\omega}$ are the vector meson decay constants,
defined by
\begin{equation}
\langle 0\vert V^{\rho ,\omega}_\mu \vert \rho ,\omega\rangle \equiv
{m^2_{\rho ,\omega }\over g_{\rho ,\omega}}\epsilon_\mu\ ,
\eqnum{2.2}\label{twopttwo}\end{equation}
and $\Pi^{\rho \omega}_{\mu\nu}$ with the off-diagonal
element of the vector meson propagator.  This amounts to assuming
that the isospin-unmixed $I=1$ $\rho$ state, $\rho^{(0)}$,
couples only to $V^\rho_\mu$, and the isospin-unmixed $I=0$ $\omega$
state, $\omega^{(0)}$, only to $V^\omega_\mu$, the isospin-breaking
contribution to $\Pi_{\mu\nu}$ of Eqn.~(2.2) from the $\rho ,\omega$
region then resulting solely from the $\rho^{(0)}$-$\omega^{(0)}$
mixing in the meson propagator.  In this interpretation, fixing
the imaginary part of the correlator in the $\rho$, $\omega$ region
(via the sum rule analysis) allows one to obtain the isospin-breaking
parameters of the imaginary part of the vector meson propagator, and,
via a dispersion relation, the behavior of the off-diagonal element of the
propagator off-shell.  However, as explained above, such a possibility
is excluded on general grounds.  The problem with going from $A_0$
and $\Pi^{\rho\omega}_{\mu\nu}$ of Eqn.~(2.2) to the interpretation
of these quantities in Ref.~\onlinecite{ref12} is that, not one, but
three sources of isospin breaking exist in the contributions to $\Pi_{\mu\nu}$
from the $\rho$, $\omega$ region:  that due to $\rho^{(0)}$-$\omega^{(0)}$
mixing (discussed above), that due to the direct coupling of
$V^\rho_\mu$ to $\omega^{(0)}$, and that due to the direct coupling
of $V^\omega_\mu$ to $\rho^{(0)}$.  The same $\Delta I=1$ strong
operator which gives rise to non-zero $\rho^{(0)}$-$\omega^{(0)}$
mixing will also necessarily give rise to the latter two couplings.
These couplings would be described by new isospin breaking parameters,
$\phi^{(\rho )\omega}$ and $\phi^{(\omega )\rho}$,
\begin{eqnarray}
&&\langle 0\vert V^\omega_\mu \vert \rho^{(0)}\rangle \equiv
{m^2_\omega\over g_\omega}\phi^{(\omega )\rho}\epsilon_\mu \nonumber \\
&&\langle 0\vert V^\rho_\mu \vert \omega\rangle \equiv
{m^2_\rho\over g_\rho}\phi^{(\rho )\omega}\epsilon_\mu
\eqnum{2.3}\label{twoptthree}
\end{eqnarray}
where $\phi^{(\omega )\rho}$, $\phi^{(\rho )\omega}$
are, in general,
$q^2$-dependent, and also interpolating-field-dependent off-shell.
Thus, off-shell, the $\rho$-$\omega$ region contribution to
$\Pi_{\mu\nu}$ depends not only on the (interpolating-field-choice-dependent)
isospin-breaking parameters of the off-diagonal element of the vector
meson propagator, but also on the (interpolating-field-choice-dependent)
isospin-breaking parameters $\phi^{(\omega )\rho}$, $\phi^{(\rho )\omega}$.
The total contribution is independent of the interpolating
field choice, but the individual contributions are not.  One is,
of course, free to choose a convenient set of $\rho$, $\omega$
interpolating fields and work with these, provided one
calculates contributions to S-matrix elements.  Since, to ${\cal O}
(m_d-m_u)$ Eqn.~(2.2) remains valid when we replace $\rho$ and
$\omega$ with $\rho^{(0)}$ and $\omega^{(0)}$, the fields
\begin{eqnarray}
&&\rho^{(0)c}_\mu\equiv {g_\rho\over m^2_\rho}V^\rho_\mu\nonumber \\
&&\omega^{(0)c}_\mu\equiv {g_\omega \over m^2_\omega }V^\omega_\mu\ ,
\eqnum{2.4}\label{twoptfour}\end{eqnarray}
satisfy$\langle 0\vert \rho^{(0)c}_\mu\vert \rho^{(0)}\rangle
=\epsilon_\mu$ and $\langle 0\vert \omega^{(0)c}_\mu\vert \omega^{(0)}
\rangle =\epsilon_\mu$, and hence
serve as possible choices of interpolating fields for
$\rho^{(0)}$ and $\omega^{(0)}$.
With this choice of interpolating fields (and not with others) one
obtains
\begin{equation}
\Pi_{\mu\nu}(q^2)={m^2_\rho\over g_\rho}{m^2_\omega\over g_\omega}
\Delta^{(c)\rho\omega}_{\mu\nu}(q^2)\eqnum{2.5}\label{twoptfive}
\end{equation}
where $\Delta^{(c)\rho\omega}_{\mu\nu}(q^2)$ is the off-diagonal
element of the vector meson propagator for the interpolating field
choice above.  If one simultaneously evaluates, e.g. $NN\rho$,
$NN\omega$ vertex form factors using the same interpolating
fields, one could, of course, piece the resulting vertices and
propagators together to obtain contributions to NN scattering
S-matrix elements which are independent of field choice.  For a
general choice of interpolating field, however, neither
$\Pi_{\mu\nu}$ nor $\Pi^{\rho\omega}_{\mu\nu}$ is proportional
to $\Delta^{\rho\omega}_{\mu\nu}$.  Given the existence of QCD
sum rules and ChPT methods, which are rather efficient at handling
current-current correlators and vector current vertex functions, the
choice (2.4) for vector meson interpolating
fields would appear to be a convenient and sensible one.
With this choice, Eqn.~(2.1) provides the basis of a spectral
representation of $\Delta^{(c)\rho\omega}$, but for other
choices of the vector meson interpolating fields this is not the case.

Note that the above discussion also clarifies one ongoing point of
debate in the literature, namely that concerning the $q^2$-dependence of
the quantity $\theta^{\rho\omega}(q^2)$ appearing in Eqn.~(1.3).
Defining $\hat{\Pi} (q^2)$ by
\begin{equation}
\Pi_{\mu\nu}(q^2)\equiv (g_{\mu\nu}-q_\mu q_\nu /q^2)\hat{\Pi}(q^2)\ ,
\eqnum{2.6}\label{twoptsix}\end{equation}
the absence of massless singularities implies that $\hat{\Pi}(0)=0$
\onlinecite{ref14}~.
This in turn implies, with
\begin{equation}
\Delta^{(c)\rho\omega}_{\mu\nu}(q^2)\equiv -(g_{\mu\nu}-q_\mu q_\nu
/q^2)\Delta^{(c)\rho\omega}(q^2)\ ,\eqnum{2.7}\label{twoptseven}
\end{equation}
$\Delta^{(c)\rho\omega}(q^2)=0$, and hence $\theta^{\rho\omega}(0)=0$.
Since this is true for one choice
of the vector meson interpolating fields, it is incumbent upon those
advocating
\begin{equation}
\theta^{\rho\omega}(q^2)=\theta^{\rho\omega}(m^2_\omega )
\eqnum{2.8}\label{twopteight}
\end{equation}
to explicitly demonstrate the existence of an interpolating field choice
for the  vector mesons
for which Eqn. (2.8) is valid; the relation
cannot be true in general.

\section{The QCD Sum Rule Analysis of $\Pi_{\mu\nu}(q^2)$ Revisited}
\label{sec:III}

With the above discussion in mind, let us turn to the sum rule analysis of
the vector correlator, first briefly reviewing the treatment and results
of Refs.~\onlinecite{ref12,ref17}~.  The sum rule approach consists of writing
an operator product expansion (OPE) representation
for the correlator, valid in the region of
validity of perturbative QCD, and a second, phenomenological,
representation in terms of hadronic parameters, and then Borel
transforming both.  The Borel transform serves to extend
the ranges of validity of both representations and, in addition,
to (1) emphasize the
operators of lowest dimension in the OPE representation and
(2) give higher weight to the parameters of the lowest lying resonances
in the phenomenological representation.
One then matches the transformed representations in order to
make predictions for the relevant hadronic parameters.

The
OPE for the correlator of interest was performed long ago\onlinecite{ref17}~.
Truncating the expansion at operators of dimension six, one finds
that, defining $\Pi (q^2)$ by
\begin{equation}
\Pi_{\mu\nu}\equiv (q_\mu q_\nu -q^2g_{\mu\nu})
\Pi (q^2)\ ,\eqnum{3.1}\label{threenew}\end{equation}
one has
\begin{equation}
\Pi^{OPE} (Q^2) = {1\over 12}\left[ -c_0\log (Q^2)
+{c_1\over Q^2} +{c_2\over Q^4}+{2c_3\over Q^6}\right]
\eqnum{3.2}\label{threeptone}
\end{equation}
where $Q^2=-q^2$ and
\begin{eqnarray}
&&c_0={\alpha_{EM}\over 16\pi^3}\nonumber \\
&&c_1={3\over 2\pi^2}(m_d^2-m_u^2)\nonumber \\
&&c_2=\left( {m_d-m_u\over m_d+m_u}\right) 2f^2_\pi m^2_\pi
\left[ 1+\left( {\gamma\over 2+\gamma}\right) \left( {m_d+m_u
\over m_d-m_u}\right)\right]\nonumber \\
&&c_3=-{224\over 81}\pi\left[ \alpha_s\langle \bar{q} q\rangle_0\right]^2
\left[ {\alpha_{EM}\over 8\alpha_s (\mu^2)} - \gamma\right]
\eqnum{3.3}\label{threepttwo}\end{eqnarray}
with $\gamma\equiv \langle\bar{d} d\rangle_0/\langle\bar{u} u\rangle_0 -1$.
Taking for the phenomenological representation (in the narrow
resonance approximation)
\begin{eqnarray}
{\rm Im}\, \Pi^{phen}(q^2)&=&
{\pi\over 12}\left[ f_\rho \delta (q^2-m_\rho^2)
-f_\omega \delta (q^2 -m_\omega^2) +f_{\rho^\prime}\delta (q^2-
m_{\rho^\prime}^2)-f_{\omega^\prime}\delta (q^2-m_{\omega^\prime}^2)\right] +
\nonumber\\
&&\qquad\qquad  {\alpha_{EM}\over 192\pi^2}\ ,\eqnum{3.4}\label{threeptthree}
\end{eqnarray}
(where $f_\rho$, $f_\omega$, $f_{\rho^\prime}$, $f_{\omega^\prime}$ may
be thought of as the parameters to be determined from the sum rule analysis)
one finds, upon Borel transformation and matching,
\begin{eqnarray}
&&{1\over M^2}\left[ f_\rho \exp (-m_\rho^2/M^2)-f_\omega\exp (-m_\omega^2/M^2)
+f_{\rho^\prime} \exp (-m_{\rho^\prime}^2/M^2)
-f_{\omega^\prime}\exp (-m_{\omega^\prime}^2/M^2)\right]\nonumber\\
&&\qquad\qquad +{\alpha_{EM}\over 16\pi^3}\exp (-s_0/M^2)
=c_0 +{c_1\over M^2} +{c_2\over M^4}+{c_3\over M^6}\ ,
\eqnum{3.5}\label{threeptfour}
\end{eqnarray}
where $M$ is the Borel mass.  As pointed out in Ref.~\onlinecite{ref12}~,
to ${\cal O}(\delta m^2,\ \delta {m^\prime}^2)$, where
$\delta m^2=m_\omega^2-m_\rho^2$, $\delta {m^\prime}^2=m_{\omega^\prime}^2
-m_{\rho^\prime}^2$, Eqn.~(3.5) can be rewritten in terms of the
parameters $\xi$, $\beta$, $\xi^\prime$ and $\beta^\prime$, where
\begin{eqnarray}
&&\xi ={\delta m^2\over m^4}\left({f_\rho +f_\omega}\over 2\right)\nonumber \\
&&\beta ={(f_\omega -f_\rho )\over m^2\, \xi}\nonumber \\
&&\xi^\prime ={\delta {m^\prime}^2\over{m^\prime}^4}
\left({f_{\rho^\prime} +f_{\omega^\prime}}\over 2\right)\nonumber \\
&&\beta^\prime ={(f_{\omega^\prime} -f_{\rho^\prime} )\over
{m^\prime}^2\, \xi^\prime}\eqnum{3.6}\label{threeptfive}\end{eqnarray}
with $m^2\equiv (m_\rho^2+m_\omega^2)/2$ and
${m^\prime}^2\equiv (m_{\rho^\prime}^2+m_{\omega^\prime}^2)/2$, as
\begin{eqnarray}
&&\xi{m^2\over M^2}\left( {m^2\over M^2}-\beta\right)\exp (-m^2/M^2)
+\xi^\prime {{m^\prime}^2\over M^2}\left( {m^\prime}^2 -\beta^\prime \right)
\exp (-{m^\prime}^2/M^2) \nonumber\\
&&\qquad +{\alpha_{EM}\over 16 \pi^3}\exp (-s_0/M^2)
=c_0 +{c_1\over M^2}+{c_2\over M^4}+{c_3\over M^6}\ .
\eqnum{3.7}\label{threeptsix}
\end{eqnarray}
If $c_{0-3}$ were precisely known, Eqn.~(3.5) or Eqn.~(3.7) could,
in principle, be used to determine the parameters $\xi$, $\beta$,
$\xi^\prime$, $\beta^\prime$.  There are, however, some uncertainties in the
values of the $c_i$, associated with the imprecision in our
knowledge of the values of the four-quark condensates and of the
isospin-breaking ratio of the $\langle \bar{u} u\rangle_0$ and
$\langle \bar{d} d\rangle_0$ condensates.  The authors of
Ref.~\onlinecite{ref12} (which updates Ref.~\onlinecite{ref17} )
consider a range of possibilities for these quantities, and
also take for $r\equiv (m_d-m_u)/(m_d+m_u)$ the value $r=0.28$, obtained
from an analysis of pseudoscalar isomultiplet splittings\onlinecite{ref24}
employing Dashen's theorem\onlinecite{ref25} for the electromagnetic
contributions to these splittings.  The last ingredient of the analysis
of Ref.~\onlinecite{ref12}
is the imposition of an external constraint on the hadronic parameter
$\xi$, based on the observed interference in the $\rho$-$\omega$
interference region in $e^+e^-\rightarrow\pi^+\pi^-$.  This
constrained value, $\xi =1.13\times 10^{-3}$, is based on
(1) the assumed connection between the correlator and the propagator
(presumably valid for the essentially on-shell value of the mixing,
though not elsewhere) and (2) the assumption that direct $\omega^{(0)}
\rightarrow\pi\pi$ contributions to $e^+e^-\rightarrow\pi^+\pi^-$
can be neglected (see Ref.~\onlinecite{ref26} for a discussion of
these issues).  There appears to be no particularly good reason for
the latter assumption, and, indeed, it would seem appropriate to
allow $\xi$ to be fit by the sum rule analysis as a test of
this assumption (as will be done below), but let us follow the analysis
of Ref.~\onlinecite{ref12} for the moment.  Using the sum rule,
Eqn.~(3.7), and imposing the constraint $\xi =1.13\times 10^{-3}$, as
discussed above, the authors of Ref.~\onlinecite{ref12} solve for
$\beta$, $\xi^\prime$ and $\beta^\prime$ for four different input
sets $\{ c_i\}$.  Using the expression (3.4) for ${\rm Im}\, \Pi^{phen}(q^2)$
and the fact that $\Pi (q^2)$ satisfies an unsubtracted dispersion
relation, one may show that, to first order in $\delta m^2$ and
$\delta {m^\prime}^2$,
\begin{equation}
{\rm Re}\, \Pi (0) ={1\over 12}\left[ \xi (1-\beta )+\xi^\prime
(1-\beta^\prime )\right]\ .\eqnum{3.8}\label{threeptseven}\end{equation}
Using the values of the parameters obtained in Ref.~\onlinecite{ref12}~,
one finds that the ratios of the contributions to ${\rm Re}\, \Pi (0)$ from
the $\rho^\prime$-$\omega^\prime$ region to those from the
$\rho$-$\omega$ region are $1.8$, $0.8$, $0.3$ and $0.8$ for
input sets I, II, III, IV, respectively.  The failure of the results
to be dominated by the nearby ($\rho$, $\omega$) singularities
suggests that the phenomenological form employed for the spectral
function may well be incomplete, either in missing low-lying contributions
or in failing to include the effect of even more distant singularities.
If we consider Eqns.~(3.4) and (3.8) for a moment
an interesting possibility becomes
evident.  If one had all isospin-breaking effects generated solely
by $\rho^{(0)}$-$\omega^{(0)}$ mixing, and if the physical vector
mesons were a simple rotation of the isospin-pure basis (not in
general true when the wavefunction renormalization matrix of
the system is non-diagonal), we would have $f_\rho =f_\omega$
for $f_\rho$, $f_\omega$ as written in Eqn.~(3.4).  While the
assumptions required to arrive at this conclusion are certainly not
satisfied in general, this nonetheless indicates that there
should be significant cancellation between the $\rho$ and $\omega$
contributions to the correlator.  Thus, a single isolated resonance,
even with a coupling much smaller than that of the $\rho$ or $\omega$,
could in fact contribute significantly to $\Pi_{\mu\nu}$.  This suggests
that the $\phi$ contribution to ${\rm Im}\, \Pi_{\mu\nu}$, neglected in
Ref.~\onlinecite{ref12}, may well be non-negligible.  In fact we
can make a rough estimate of the expected size of $f_\phi$ (where
$f_\phi$ is defined by adding a contribution ${\pi\over 12} f_\phi \delta (q^2
-m_\phi^2)$ to ${\rm Im}\, \Pi^{phen}(q^2)$ in Eqn.~(3.4)) as follows.
$\phi$ is known to be not quite pure $\bar{s} s$.  If, e.g., we
take the Particle Data Group (PDG)\onlinecite{ref27} value for
the octet-singlet mixing angle, $\theta =39^0$ (quadratic fit),
$\phi \simeq \phi^{(0)} -\delta \omega^{(0)}$, where $\phi^{(0)}$
is the pure $\bar{s} s$ state and $\delta =.065$ rad is
the deviation of $\theta$ from ideal mixing.  The contribution of
the $\phi$ pole term to $\Pi_{\mu\nu}$ due to mixing in
the propagator should then be of order $-\delta$ times that
associated with the $\omega$ pole, i.e. $\simeq 0.065\, f_\omega
\simeq 0.065\, f_\rho$.  There will, of course, also, in general,
be isospin-breaking contributions from direct couplings to the
current vertices, not just from mixing in the propagator, but
the above discussion shows that $f_\phi\simeq (0.05-0.10)\, f_{\rho ,\omega}$
should be a reasonable
expectation.  As we will see below, this (rather crude) estimate
is indeed borne out by the sum rule analysis.

Let us, therefore, add a term ${\pi\over 12} f_\phi\delta (q^2 -m_\phi^2)$ to
${\rm Im}\, \Pi^{phen}(q^2)$ on the RHS of Eqn.~(3.4), and perform
a reanalysis of that equation.  We will follow Ref.\onlinecite{ref12}
in choosing the range of input values for the $\{ c_i\}$, with,
however, the following modifications.  First, the small
$c_1$ term dropped in Ref.~\onlinecite{ref12} will be
retained, though, as pointed out there,
it in fact has little effect on the final results.  The numerical value
is obtained by using
$(m_d+m_u)(1 {\rm GeV}) =12.5\pm 2.5 {\rm MeV}$ from Ref.~\onlinecite{ref28}
and the updated value of $r$ discussed below.  The
main modification to the input is in the parameter $r$.  There is
now considerable evidence that Dashen's theorem is significantly
violated\onlinecite{ref29,ref30,ref31}, Refs.~\onlinecite{ref30,ref31}
in particular suggesting that
\begin{equation}
(m_{K^+}^2-m_{K^0}^2)_{EM}\simeq 1.9\, (m_{\pi^+}^2-m_{\pi^0}^2)_{expt}
\eqnum{3.9}\label{threepteight}\end{equation}
(where the factor $1.9$ on the RHS of Eqn.~(3.9) is absent in Dashen's
theorem).  Using Eqn.~(3.9) in place of Dashen's theorem for the
electromagnetic contribution to the kaon mass splitting produces a
rescaling of $r$ by $1.22$.  The resulting change
in the $c_i$ is essentially to rescale the values of $c_2$ in
Ref.~\onlinecite{ref12} by this same factor.  In assessing the effect
of the uncertainties in the values of the $\{ c_i\}$ for
a given input set, the input errors on $c_2$ have also been
rescaled by this factor of $1.22$.  Finally, since the masses of
all the resonances appearing above, including the $\rho^\prime$
and $\omega^\prime$, are  known, we may take these as input and
use the sum rule to extract the isospin-breaking parameters,
$\{ f_k\}$, where $i=1\cdots 5$ correspond to $\rho$, $\omega$, $\rho^\prime$,
$\omega^\prime$ and $\phi$, respectively.  Note that, in taking
this approach, we are abandoning the constraint on $\xi$ employed in
Ref.~\onlinecite{ref12}.  If the direct $\omega^{(0)}\rightarrow\pi^+\pi^-$
coupling is, indeed, negligible in $e^+e^-\rightarrow\pi^+\pi^-$, this
will manifest itself by the value of $\xi$ resulting from the sum rule
analysis being near $1.13\times 10^{-3}$.

The analysis of the modified version of the
sum rule, (3.5), proceeds as follows.  First, from the
terms of ${\cal O}(M^0)$, $c_0=\alpha_{EM}/16\pi^3$.  One may
check that, as in Ref.~\onlinecite{ref12}~, the analysis is
very insensitive to the value of the EM threshold parameter, $s_0$.
We will, therefore, quote all results below for the value, $s_0=1.8$ GeV,
employed in a number of the results quoted in Ref.~\onlinecite{ref12}~.
Second, again as in Ref.~\onlinecite{ref12}~, we impose
the local duality relation
\begin{equation}
\int ^\infty_0 \ ds\ {\rm Im}\, \Pi^{phen}(s) ={\cal O}(\alpha_{EM},
m_q^2)\eqnum{3.10}\label{threeptnine}\end{equation}
(which is equivalent to matching the coefficients of the ${\cal O}(1/M^2)$
terms in Eqn.~(3.5)).  With the index $k=1,\cdots ,5$ labelling
$\rho$, $\omega$, $\rho^\prime$, $\omega^\prime$ and $\phi$, respectively,
as above, this relation is
\begin{equation}
\sum_k (-1)^{k+1}f_k =c_0s_0 +c_1~.
\eqnum{3.11}\label{threeptten}\end{equation}
(Note that the $c_i$ tabulated in Ref.~\onlinecite{ref12} have had
the appropriate factors of $m^2$ required to leave the
remaining coefficient dimensionless
factored out of them.  Thus, e.g., $c_1$ in Eqn.~(3.11)
is $m^2$ times that tabulated in Ref.~\onlinecite{ref12}~.)  The remaining
four relations required to obtain a solution for the five unknowns, $\{ f_k\}$,
are obtained by acting on Eqn.~(3.5) with $(-1)^n{\partial^n\over\partial
(1/M^2)^n}$ for $n=1,\cdots ,4$.  One may check that the results are
not sensitive to using precisely the PDG values for the $\rho^\prime$ and
$\omega^\prime$ masses.  Indeed, shifting either mass by $50$ MeV
induces changes of $<4\%$ in $\xi$, $<2.5\%$ in $\beta$, $<5\%$ in
$\beta^\prime$ and $<20\%$ in $\xi^\prime$.  The resulting changes
in the correlator itself are even smaller:  e.g. $\Pi (0)$ and
${d\Pi \over dq^2}(0)$ are changed by $<2\%$ by the above mass shifts.

In Table 1, the results of the modified
sum rule analysis are displayed for the
input sets I, III, IV of Ref.~\onlinecite{ref12}~, modified as
described above.
The errors shown in the table correspond
to the uncertainties in the input parameters, $c_2$ and $c_3$,
(those quoted in
Ref.~\onlinecite{ref12} in the case of
$c_3$ and the rescaled version thereof in the case of $c_2$).
The stability of the analysis is illustrated, for input set IV, in
Figs.~1-5, which display the parameters
$\xi$, $\beta$, $\xi^\prime$, $\beta^\prime$, $f_\phi$
as a function of the Borel mass, $M$, in the range $1$-$10$ GeV
(the choice of the first
four parameters, rather than corresponding $f_k$ values, is made in order
to facilitate comparison with Ref.~\onlinecite{ref12}~).
Set I generates results of comparable stability, while the results of
set III are even more stable than those of set IV.
In all three cases
a wide stability window exists in the Borel mass for all five output
parameters.
This stability window, moreover, occurs without the necessity of using
unphysical values for the the average of the $\rho^\prime$
and $\omega^\prime$ masses.
As noted previously in Ref.~\onlinecite{ref12}~, results for input set II
are considerably less stable than for the other sets:
in fact, no stability window exists anywhere in the range
$M=1$ and $M=10$ GeV, apart from for the very lower edge of the error band for
the magnitude of $c_3$, for which values input set II is very close
to the upper end of the
corresponding error band for input set I.  The instability of the
analysis for input set II
is illustrated (for the central values of $c_2$ and $c_3$)
in Fig.~6, where the parameter, $f_\phi$, is plotted
as a function of the Borel mass, $M$.  As a result of this instability,
results corresponding to input set II
are not quoted in the table; for most of the input
range (i.e. for larger values of the magnitude of $c_3$) the input set
appears, from the sum rule analysis, to be unphysical.

A number of features are evident from the results of the above analysis.
First, from Table~1, we see that the magnitude of $\xi$ differs
significantly from that which would be expected from the analysis
of $e^+e^-\rightarrow \pi^+\pi^-$, neglecting
$\omega^{(0)}\rightarrow\pi^+\pi^-$ contributions,
suggesting that the latter are, indeed,
not negligible.  It should be stressed that the
errors quoted in the table correspond to varying $c_2$ and $c_3$
separately within the range of quoted errors, and taking the
maximum variation of the resulting output.  One can obtain even
lower values of $\xi$, i.e. closer to that expected if one can
indeed neglect $\omega^{(0)}\rightarrow\pi^+\pi^-$ contributions
to $e^+e^-\rightarrow \pi^+\pi^-$, by letting $c_2$ lie at the
bottom of its error band and, simultaneously, the magnitude of
$c_3$ lie at the top of its error band in set I.  However, such
a combination (which produces $\xi =1.43\times 10^{-3}$) is quite
unstable, the values of $\xi^\prime$, e.g., varying by more than
$20\%$ between $M=3$ and $5$ GeV.  A similar result, $\xi =1.48\times 10^{-3}$,
can be obtained from set II for the central value of $c_2$ and the
lower edge of the error band for the magnitude of $c_3$, with
comparable ($\simeq 20\%$ over the range $M=3$ to $5$ GeV) instability.
All other portions of
the set II error band are even more unstable.
Thus it appears very clear that the value
$\xi =1.13\times 10^{-3}$ is excluded by the sum rule analysis.
The second observation is that the inclusion of the $\phi$ pole term in
the phenomenological representation of the correlator rectifies
the problem of the strength of the distant singularities.  This can
be seen from the relative size of $\xi$ and $\xi^\prime$ in Table 1,
but is more evident in Table 2, where the output
values for the parameters $\{ f_k\}$ are tabulated,
for the central values of the
input parameters $\{ c_i\}$, for input sets I, III, IV.  The ratios
of $f_\phi$ to $f_\omega$ are $0.062$, $0.068$ and $0.066$ for
sets I, III and IV, respectively.  This is in (better than should
be expected) agreement with the rough estimate given above, confirming
the physical plausibility of the solutions obtained.  Moreover,
$f_{\rho^\prime}$ and $f_{\omega^\prime}$ are now a factor of
$40$-$60$ smaller than $f_\rho$ and $f_\omega$.  The structure of
the resulting contributions to the correlator near $q^2=0$ is shown in Table 3,
where the $\rho$, $\omega$, and also the $\rho^\prime$,
$\omega^\prime$ contributions have been combined.  Note that the individual
$\rho$ and $\omega$ contributions are a factor of $\simeq 13$ larger
than the $\phi$ contribution, but the cancellation between them
is such that the $\phi$ contribution is approximately twice
as large as their sum.  The $\rho^\prime$-$\omega^\prime$ region
contribution is then less than $10\%$ of the $\phi$ contribution.
The more distant singularities, thus, have only a small effect, justifying,
{\it a posteriori}, the neglect of yet more distant singularities in the
phenomenological side of the sum rule analysis.  Given that
the results satisfy all the above tests for being physically
sensible and stable, it appears that the resulting values for
the correlator and its slope with respect to $q^2$ at $q^2=0$
should be taken as good estimates, within the uncertainties
resulting from those in the input parameters.  The fact that,
due to cancellation between the otherwise dominant $\rho$ and $\omega$
contributions, the $\phi$ contribution is actually dominant, no doubt
accounts for the unphysical behavior of the spectral distribution
of the correlator obtained in the absence of the $\phi$ term.  Note
that, despite the significant changes in the fit, as compared to
Ref.~\onlinecite{ref12}~, the slope of the correlator remains
large in the present results.

\section{The Correlator to One-Loop Order in ChPT}\label{sec:IV}

The starting point for the computation of the mixed-isospin correlator,
$\Pi_{\mu\nu}(q^2)$, is the effective chiral Lagrangian
of Ref.\ \onlinecite{ref21},
\begin{eqnarray}
{\cal L}_{eff} &=& {1\over 4}f^2 \text{Tr}(D_\mu\Sigma D^\mu
        \Sigma^\dagger )
        +{1\over 2}f^2 \text{Tr}[B_0 M(\Sigma +\Sigma^\dagger )]
        +L_1 \bigl[ \text{Tr}(D_\mu\Sigma D^\mu\Sigma^\dagger )
        \bigr]^2  \nonumber \\
     && \mbox{} + L_2 \text{Tr}(D_\mu\Sigma D_\nu\Sigma^\dagger )
        \text{Tr}(D^\mu\Sigma D^\nu\Sigma^\dagger )
        + L_3 \text{Tr}(D_\mu\Sigma^\dagger D^\mu\Sigma
        D_\nu\Sigma^\dagger D^\nu\Sigma)  \nonumber \\
     && \mbox{} + L_4 \text{Tr}(D_\mu\Sigma D^\mu\Sigma^\dagger )
        \text{Tr}[2B_0 M(\Sigma +\Sigma^\dagger )]
        + L_5 \text{Tr}\bigl[ 2B_0 (M\Sigma +\Sigma^\dagger M)
        D_\mu\Sigma^\dagger D^\mu\Sigma \bigr] \nonumber \\
     && \mbox{} + L_6 \bigl[ \text{Tr}[2B_0
         M(\Sigma +\Sigma^\dagger )]\bigr]^2
        + L_7 \bigl[ \text{Tr}[2B_0 M(\Sigma -\Sigma^\dagger )]\bigr]^2
         \nonumber \\
     && \mbox{}+L_8 \text{Tr}[4B_0^2(M\Sigma M\Sigma +
         M\Sigma^\dagger M\Sigma^\dagger )]
         -iL_9\text{Tr}[F_{\mu\nu}^R D^\mu\Sigma D^\nu\Sigma^\dagger
        +F_{\mu\nu}^L D^\mu\Sigma^\dagger D^\nu\Sigma ] \nonumber \\
     && \mbox{} +L_{10}\text{Tr}[\Sigma^\dagger F_{\mu\nu}^R\Sigma
         F^{L\mu\nu}] +H_1\text{Tr}[F_{\mu\nu}^RF^{L\mu\nu}+
         F_{\mu\nu}^LF^{L\mu\nu}] +H_2\text{Tr}[4B_0^2 M^2]~.
         \eqnum{4.1} \label{fourptone}
\end{eqnarray}
In Eqn.~(4.1), $B_0$ is a mass scale
related to the value of the quark condensate in the chiral limit,
$\Sigma = \text{exp}(i\vec \lambda \cdot \vec \pi /f) $
(with $\vec\lambda$ the usual $SU(3)$ Gell-Mann matrices and
$\vec\pi$ the octet of pseudoscalar (pseudo-) Goldstone boson fields),
$f$ is a dimensionful
constant, equal to $f_\pi$ in leading order, $M$ is the current
quark mass matrix, and $D_\mu$ is the covariant derivative
\begin{equation}
  D_\mu\Sigma = \partial_\mu\Sigma -i(v_\mu +a_\mu )\Sigma
                 +i\Sigma (v_\mu - a_\mu ).\eqnum{4.2} \label{fourpttwo}
\end{equation}
In Eqn.~(4.1) the external pseudoscalar sources
(which occur in the most general form of ${\cal L}_{eff}$) have already been
set to zero, and the external scalar source to $2B_0$ times
the current quark mass
matrix, since we are interested here only in the vector current correlator and,
therefore, require only the external vector sources.
For this same reason we may drop the external axial vector sources
from the expression for the covariant derivative in Eqn.~(4.2).
The left and right external source field strength tensors,
$F_{\mu\nu}^{L,R}$, then both reduce to
$F_{\mu\nu}^{L,R}=\partial_\mu v_\nu
-\partial_\nu v_\mu - i[v_\mu ,v_\nu ]$,
where  $v_\mu = {\lambda^a\over 2}v^a_\mu$, with $v_\mu^a$ the octet of
external $SU(3)$ vector fields.
In principle, Eqn.~(4.1) should be supplemented by
terms involving $Tr(F_{\mu\nu})$ in order to treat the case at hand,
since the current $V^\omega_\nu$ contains both octet and singlet
pieces.  However, to one-loop order, the additional terms do not
contribute to the isospin-mixed correlator (the correlator is
identical to $\Pi^{38}_{\mu\nu}(q^2)/3\sqrt{3}$, with
$\Pi^{38}_{\mu\nu}(q^2)=\, i\int d^4x \langle 0\vert T(V^3_\mu (x)
V^8_\nu (0))\vert 0\rangle $, to this order), so we will not
explicitly display these terms.
The unrenormalized higher order
coefficients $L_1,\cdots , L_{10}$ and $H_1,\ H_2$ appearing in Eqn.~(4.1)
contain divergent pieces which cancel those of the one-loop
graphs involving vertices from the first two terms in ${\cal L}_{eff}$
above, and also finite, renormalization-scale-dependent pieces,
$L_i^r$.  Expressions for the divergent pieces of the $L_i$, $H_i$,
relevant to one-loop calculations,
may be found in Ref.\ \onlinecite{ref21}.

Contributions to the correlator resulting from Eqn.~(4.1) are of two
types, corresponding to the two types of contribution to the
low-energy representation of the product of currents,
$V_\mu^\rho V_\nu^\omega$:  (1) those terms arising from the product of the
low-energy representations of the individual currents, $V^\rho_\mu$
and $V^\omega_\nu$ (obtained from the terms in ${\cal L}_{eff}$
linear in the $v^a_\alpha$, $a=0,\cdots ,8$), and (2) contact
terms
(generated by the terms in ${\cal L}_{eff}$ quadratic in the
$v^a_\alpha$).
To leading order (i.e. keeping
only the first two terms in ${\cal L}_{eff}$) the correlator vanishes.
This is because it is isospin-breaking and the only isospin-breaking at
leading order lies in the term involving the quark mass matrix, which
does not contain the external vector sources, and hence does not
contribute to the correlator in zero-loop graphs.  The leading
contributions to the correlator are, therefore, next-to-leading
order in the usual chiral counting.  As such, the contributions
consist of one-loop contact and non-contact graphs (where the
current vertices are obtained from the first term in the effective
Lagrangian above) and meson-field-independent contact terms from the
remainder of ${\cal L}_{eff}$.  These latter contributions, which
would in general produce terms involving the $L_i^r$, may be easily
shown to vanish for the case at hand.  Thus only the contact and non-contact
graphs mentioned above contribute.  It is straightforward to demonstrate
then that, to one-loop order, the ${\cal O}(m_d-m_u)$
expression for the correlator is
\begin{equation}
\Pi (q^2)={1\over 12}\left[ {\log (m_{K^0}^2/m_{K^+}^2)\over 48\pi^2}
+\left( {4m_{K^0}^2\over 3q^2}-{1\over 3}\right) \bar{J}_{K^0}(q^2)
-\left( {4m_{K^+}^2\over 3q^2}-{1\over 3}\right) \bar{J}_{K^+}(q^2)\right]
\eqnum{4.3}\label{fourptthree}\end{equation}
where
\begin{equation}
\bar{J}_P(q^2)=-{1\over 16\pi^2}\int_0^1 \, dx\ \log \left[ 1-x(1-x)q^2/m_P^2
\right]~.\eqnum{4.4}\label{fourptfour}\end{equation}
For our purposes we will not need the general expression for $\bar{J}$
(which is quoted in Appendix A of Ref.~\onlinecite{ref21}~), but only
the behavior near $q^2=0$, which is given by
\begin{equation}
\bar{J}_P(q^2)={1\over 96\pi^2}{q^2\over m_P^2}+{1\over 960\pi^2}{q^4\over
m_P^4}+\cdots \ .\eqnum{4.5}\label{fourptfive}\end{equation}
In Eqn.~(4.3), $m^2_{K^0,K^+}$ are the leading-order expressions for the
kaon squared-masses, $m_{K^0}^2=B_0(m_s+m_d)$ and $m_{K^+}^2=B_0(m_s+m_u)$
and terms have been kept only to ${\cal O}(m_d-m_u)$.  As such, we
must also expand all terms occuring there to the same order.  Doing so, and
making the expansion of Eqn.~(4.5) for the loop integrals $\bar{J}_{K^0,
K^+}$, we obtain, for the behavior of the correlator in the vicinity of
$q^2=0$,
\begin{equation}
\Pi (q^2)={1\over 12}\left( {(m_{K^0}^2-m_{K^+}^2)\over
48\pi^2 \bar{m}_K^2}\right)
\left( 1+{q^2\over 10\bar{m}_K^2}+\cdots \right)~,\eqnum{4.6}
\label{fourptsix}\end{equation}
where $\bar{m}_K^2$ is the average of the $K^+$ and $K^0$ squared masses.
Thus, $12\Pi (0)=(m_{K^0}^2-m_{K^+}^2)/ 48\pi^2 \bar{m}_K^2$,
where the kaon mass difference is that due to the strong isospin-breaking,
i.e., with the electromagnetic contribution removed.  Using Eqn.~(3.9)
for the electromagnetic contribution, we find that the RHS of this
expression is $5.5\times 10^{-5}$, to be compared with the results of
the sum rule analysis, $\simeq 1\times 10^{-3}$.  The one-loop ChPT
result is a factor of $\simeq 20$ smaller than the sum rule result.

The discrepancy between the one-loop ChPT and sum rule analyses for the
correlator near $q^2=0$ should actually not come as a complete surprise.
Indeed, when the leading-order contribution to a physical quantity (order
2 in the chiral expansion) vanishes, as it does here, one has no
obvious scale to use in judging whether or not the next-to-leading-order
contribution obtained is abnormally small, i.e., whether or not the
resulting one-loop expression is likely to represent a well-converged
approximation to the whole chiral expansion.  In fact, the structure
of the expression, (4.3), above for the correlator, $\Pi (q^2)$,
is such as to suggest that it is unlikely to be well-converged.  The reason
for this statement is that Eqn.~(4.3) is independendent of the
low-energy constants (LEC's), $L_i^r$, and results purely from
one-loop graphs involving internal kaon loops.  Such loops, for the
non-contact graphs, are well-known to be suppressed in size (the
coefficient of $q^2$ in the leading term of $\bar{J}_K$ in
Eqn.~(4.5), e.g., is a factor of $m_\pi^2/m_K^2$ smaller than for
the corresponding $\pi$ loop integral, $\bar{J}_\pi$) and, moreover,
in the case at hand, i.e. the correlator $\Pi (q^2)$,
those terms in which this suppression would be lifted by the
presence of the $m_K^2/q^2$ factor in the coefficient multiplying
$\bar{J}_K(q^2)$ cancel, since the expression for the
correlator involves the difference of the $K^+$ and $K^0$ loop
contributions.  The correlator, of course, has a cut beginning
at $q^2=4m_\pi^2$, associated with $\pi\pi$ intermediate states,
but such intermediate states do not enter until two-loop order in
the chiral expansion.  Since the relevant $\pi$ loop integral is
intrinsically much larger than its kaonic counterpart, it is likely
that the two-loop contributions will not be negligible, despite being
higher order in the chiral expansion.

Other examples of slow convergence of the chiral series when the leading
contribution vanishes and the next-to-leading order contribution results
purely from loop graphs (i.e. is independent of the fourth-order LEC's,
$L_i^r$) are, in fact, already known.  One is the process $\gamma\gamma
\rightarrow\pi^0\pi^0$, whose amplitude, to one-loop order, receives
contributions only from loop graphs (though in this case, loop graphs
with internal $\pi$ lines).  The one-loop
expression\onlinecite{ref33,ref34}
deviates from the experimental amplitude\onlinecite{refexp1} even very close to
threshold, and one finds that extending the
calculation to two-loop order (sixth order in the chiral expansion)
produces corrections to the one-loop result of order
$30\%$\onlinecite{ref35}~, which corrections
bring the amplitude into agreement with experiment.  Even more
closely similar to the case at hand is the process $\eta\rightarrow
\pi^0\gamma\gamma$.  The one-loop amplitude again has no leading
term and no contributions from the fourth order LEC's, but here,
although there are $\pi$ loop contributions, these contributions
are suppressed by a factor $(m_d-m_u)$.  The $K$ loop contributions
are naturally small, as noted above.  The result is that the one-loop
prediction for the partial rate \onlinecite{ref36} is a factor of
$\simeq 170$ smaller than observed experimentally\onlinecite{ref27}~.

It is worth considering the process $\eta\rightarrow\pi^0\gamma\gamma$
in somewhat more detail since, not only does it closely parallel
the case at hand, but the physical origin of the smallness of the
one-loop result for this process is well-understood.  The
source of the problem lies in the fact
that the dominant contribution to the amplitude is known to be due to vector
meson exchange \onlinecite{ref37,ref38}~.  As is well-known
\onlinecite{ref20,ref39,ref40}~, it is possible to make standard field choice
for the various meson resonances and write an effective chiral
Lagrangian which includes both these resonances and the octet
of pseudoscalar (pseudo-) Goldstone bosons.  One may then integrate out the
(heavy) resonance fields to obtain an effective Lagrangian of
the form ${\cal L}_{eff}$ for the pseudoscalars alone.  The
resonance contributions to the LEC's
are then determined by the coupling parameters of the original,
extended Lagrangian (which are fixed by experimental data).  The
effect of the resonances (for the initial field choices used) then
lies solely in their contributions to the $L_i^r$\onlinecite{ref39,ref40}~.
Those $L_i^r$ to which the vector and axial-vector mesons can
contribute ($L_i^r$, $i=1,2,3,9,10$) are known to be essentially
saturated by these contributions\onlinecite{ref39,ref40}~.  Thus,
the absence of any $L_i^r$-dependence in the one-loop amplitude
implies the absence of the effects of vector meson exchange
(for the given initial choice of vector meson interpolating fields) and if,
as seems to be the case for $\eta\rightarrow\pi^0\gamma\gamma$,
vector meson exchange is the dominant contribution, the one-loop
amplitude can be expected to be a
poor representation of the full chiral series.
The vector meson contributions, in this case, first appear as
tree-level contributions arising from the ${\cal O}(p^6)$ part of
the effective Lagrangian, not included in Eqn.~(4.1) above
(the general form of the ${\cal O}(p^6)$ part of the effective
Lagrangian is given in Ref.~\onlinecite{ref41}~).  Thus, only by
including these contributions (which requires, for consistency,
a full two-loop-order calculation) can one hope to obtain a
well-converged approximation to the full chiral series for
the amplitude.

The $\eta\rightarrow\pi^0\gamma\gamma$
discussion above can obviously be transferred directly to the
case of the mixed-isospin vector current correlator under
consideration here.  We expect significant contributions to the
correlator from the vector meson resonances and, for a particular
choice of vector meson interpolating fields, these contributions
are completely absent from the one-loop result.  As a consequence,
we can expect significant contributions from the tree-level
${\cal O}(p^6)$ terms in which such contributions reside.  As
already discussed, the ${\cal O}(p^6)$ loop contributions (arising from
two-loop graphs with lowest-order vertices and one-loop graphs
with a single ${\cal O}(p^4)$ vertex) may also be significant.
A two-loop calculation is, therefore, almost certainly required
in order to obtain convergence of the chiral series for the
mixed-isospin correlator.  Similar statements hold for the
related correlator, $\Pi^{38}_{\mu\nu}(q^2)$, for which
work on the two-loop calculation is in progress\onlinecite{ref42}~.
Note that, in the latter case, only a single combination of the ${\cal O}(p^6)$
LEC's enters the two-loop result.  This combination, which in
the notation of Ref.~\onlinecite{ref43} (where the analogous
$\Pi^{33}_{\mu\nu}$ and $\Pi^{88}_{\mu\nu}$ correlators are
computed to two-loop order), is written
$Q^0(\mu )-3L^{(-1)}_9(\mu )-3L^{(-1)}_{10}(\mu )$,
with $\mu$ the renormalization scale,
is in principle obtainable from experimental data using the chiral
sum rules of Ref.~\onlinecite{ref43} (Eqns.~(97) and (98) therein).
In the case at hand, one further ${\cal O}(p^4)$ LEC and one further
${\cal O}(p^6)$ LEC
will be present
at the two-loop level, but
the sum rule analysis above, in combination with a full two-loop
evaluation of $\Pi (q^2)$, would provide a useful constraint on
these parameters, albeit it with the $\simeq 20-30\%$
errors displayed in Table III and associated with the uncertainties
in the values of the input parameters $\{ c_i\}$ which determine
the correlator near $q^2=0$.  A similar sum rule analysis of the
correlator $\Pi^{38}_{\mu\nu}$ would constrain the combination
$Q^0(\mu )-3L^{(-1)}_9(\mu )-3L^{(-1)}_{10}(\mu )$, mentioned above.

It is interesting to note that the relation between the sum rule and
ChPT results for the mixed-isospin vector correlator is effectively
the reverse of what occurs in the mixed-isospin axial correlator
case.  In the latter case, the ChPT\onlinecite{ref16}
and sum rule\onlinecite{ref13} results for
the value of that piece of the correlator proportional to $q_\mu q_\nu$
at $q^2=0$ are comparable, but the ChPT result for the slope of
the correlator with $q^2$ is more than an order of magnitude larger
than that obtained from a sum rule analysis analogous to that
employed above for the vector correlator case\onlinecite{ref16}~.
The source of the discrepancy, in the axial correlator case, is that the
sum rule result for the slope has the incorrect chiral behavior,
being in fact missing its leading contribution in the chiral expansion.  This
problem with the sum rule treatment is easily exposed using chiral
methods, but is completely non-obvious without them.  In the
present case, since we do not know what portion of the vector
meson masses survives in the chiral limit, we cannot make
as precise statements about the required chiral behavior of the
vector correlator.  There is, however, no obvious problem with
the form of the sum rule result above.  The sum rule result, moreover,
provides clear evidence to indicate that the chiral series for
the vector correlator is indeed, as suggested by analogy to
the known behavior of the $\eta\rightarrow\pi^0\gamma\gamma$
process, slowly converging.  The sum rule result, in this case,
should also provide useful input for the two-loop analysis in
ChPT.  The two examples clearly indicate the advantages of
applying {\it both} methods, within their common range of
validity, in any given physical process.

\section{Summary of results}\label{sec:level1}

The basic results of the paper are as follows.  We have demonstrated
that (1) in making a sum rule analysis of the mixed-isospin vector
current correlator, it is necessary to include the $\phi$ pole
term in the phenomenological form of the representation of the
correlator, and that, when one does so, the spectral structure of
the correlator becomes physically sensible; (2) the expression
for the correlator away from $q^2=m_\omega^2$ has no general
interpretation as the off-diagonal element of a vector-meson
propagator except for a particular vector meson interpolating
field choice; (3) the freedom of field redefinition shows that
the isospin-breaking factor $\theta^{\rho\omega} (q^2)$, which occurs in
the numerator of the expression, (1.3), for the off-diagonal element
of the vector meson propagator, cannot, in general, be taken to be
independent of $q^2$;
(4) the behavior of the correlator near $q^2=m_\omega^2$
suggests that the direct $\omega^{(0)}\rightarrow\pi^+\pi^-$
contribution to $e^+e^-\rightarrow\pi^+\pi^-$ is not negligible
in the $\rho$-$\omega$ interference region; (5) the discrepancy
between the behavior of the correlator near $q^2=0$ as obtained
from the sum rule analysis and from ChPT to one-loop indicates a
slow convergence of the chiral series for the correlator and,
in consequence, the necessity of a two-loop calculation of this
quantity in ChPT.  The sum rule result for the correlator near $q^2=0$
can then be used, in such a calculation, to constrain the ${\cal O}(p^6)$
LEC's of ChPT.

\acknowledgements

The hospitality of the Department of Physics and Mathematical Physics of
the University of Adelaide and the continuing financial support of the
Natural Sciences and Research Engineering Council of Canada are gratefully
acknowledged. The author also wishes to thank Tony Williams,
Tony Thomas, Terry Goldman,
and Jerry Stephenson for numerous useful discussions on the issues of
charge symmetry breaking in few-body systems.

%
%
\begin{figure} [htb]
\centering{\
\epsfig{angle=0,figure=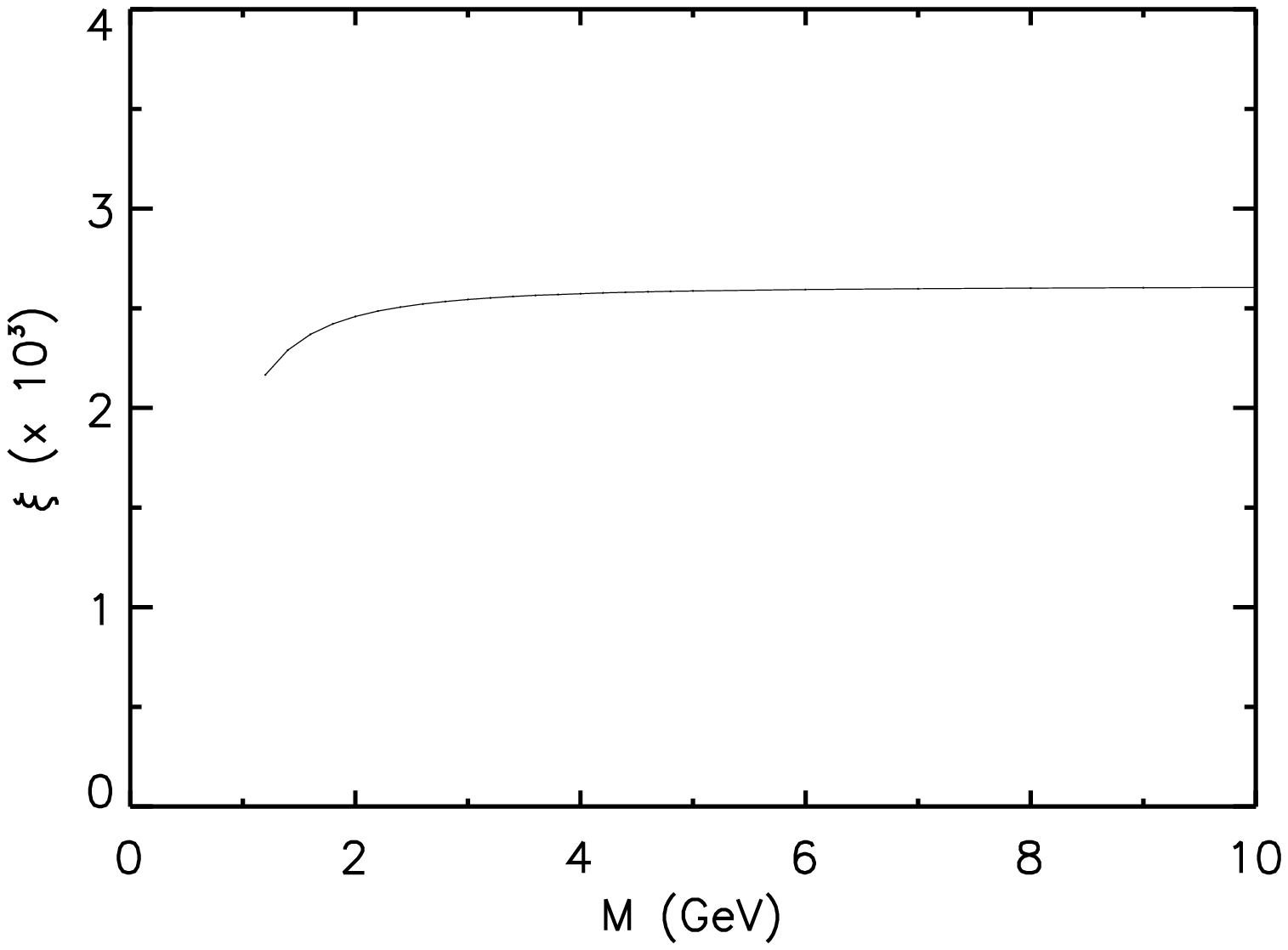,height=9cm}  }
\caption{Dependence of $\xi$ on the Borel mass, $M$, for modified
input set IV.}
\label{one}
\end{figure}

\begin{figure} [htb]
\centering{\
\epsfig{angle=0,figure=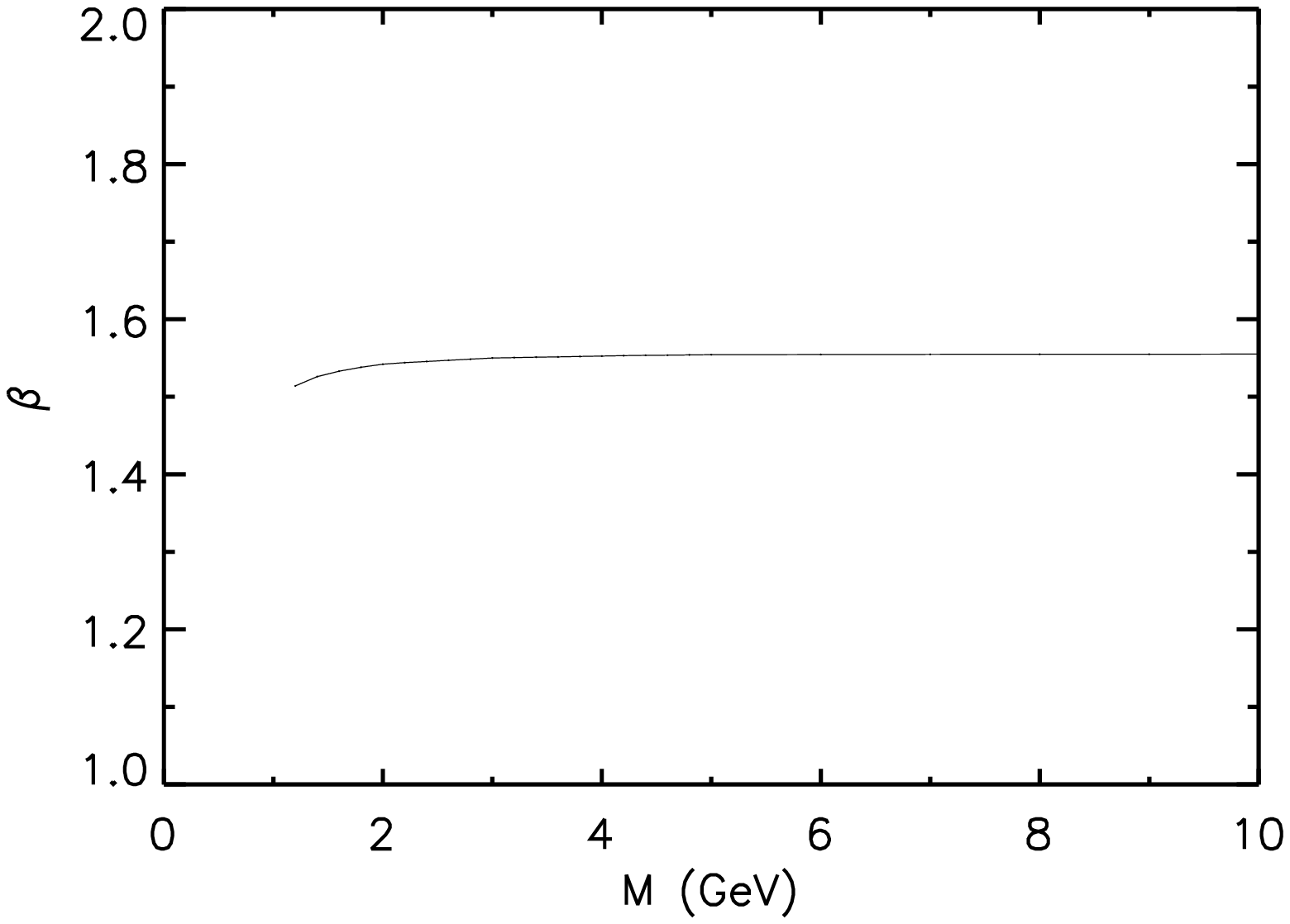,height=9cm}  }
\caption{Dependence of $\beta$ on the Borel mass, $M$, for modified
input set IV.}
\label{two}
\end{figure}

\begin{figure} [htb]
\centering{\
\epsfig{angle=0,figure=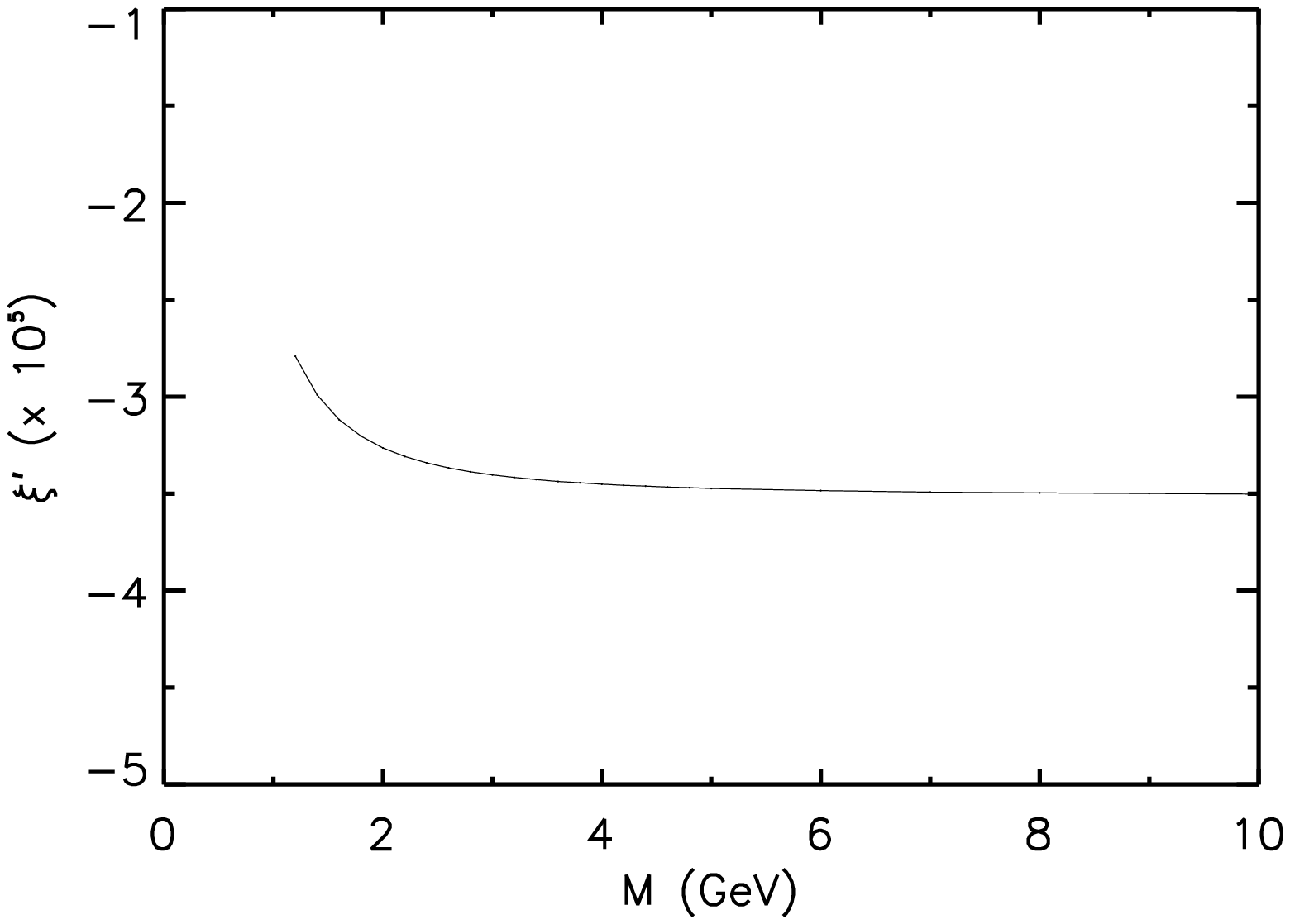,height=9cm}  }
\caption{Dependence of $\xi^\prime$ on the Borel mass, $M$, for modified
input set IV.}
\label{three}
\end{figure}

\begin{figure} [htb]
\centering{\
\epsfig{angle=0,figure=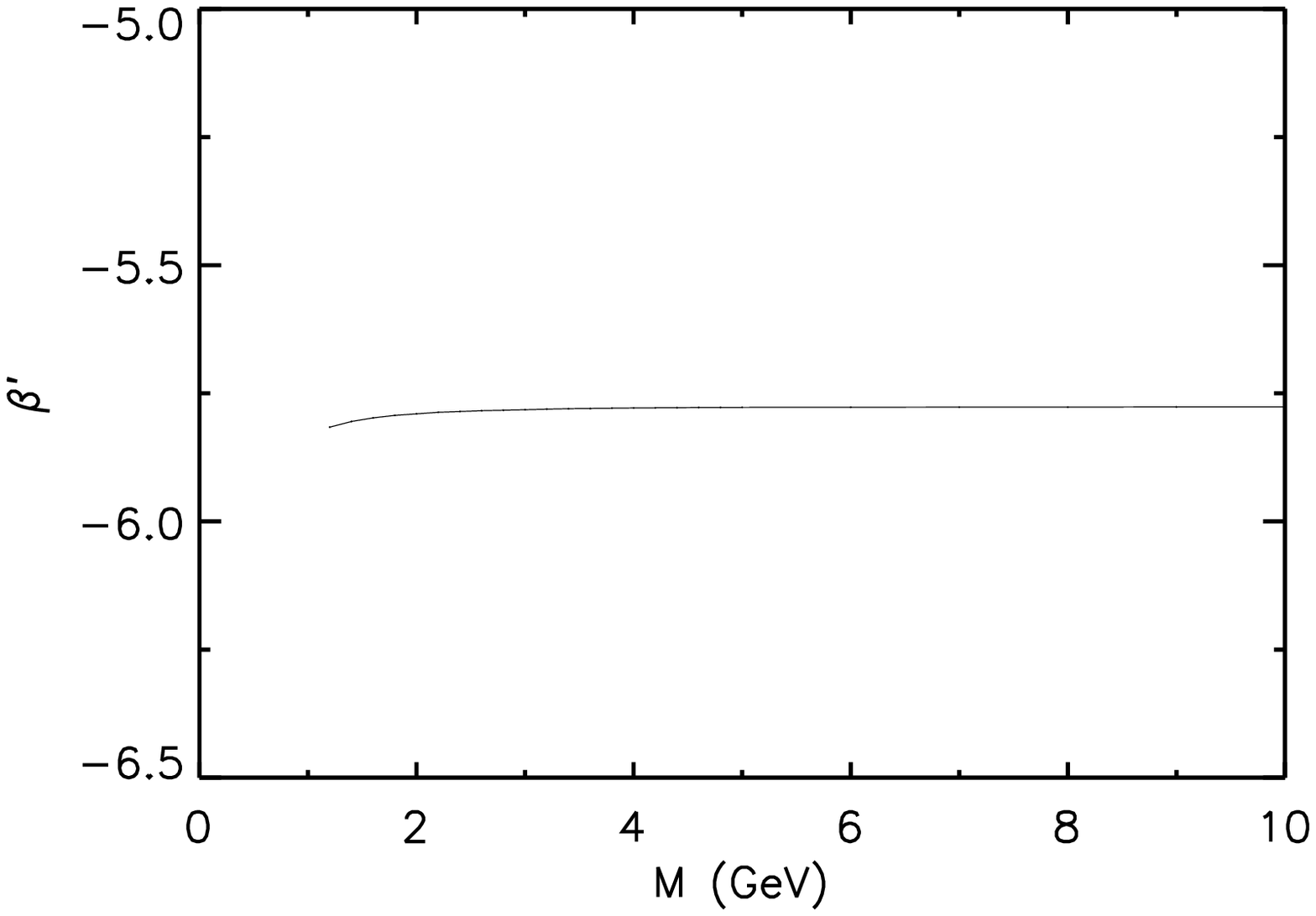,height=9cm}  }
\caption{Dependence of $\beta^\prime$ on the Borel mass, $M$, for modified
input set IV.}
\label{four}
\end{figure}

\begin{figure} [htb]
\centering{\
\epsfig{angle=0,figure=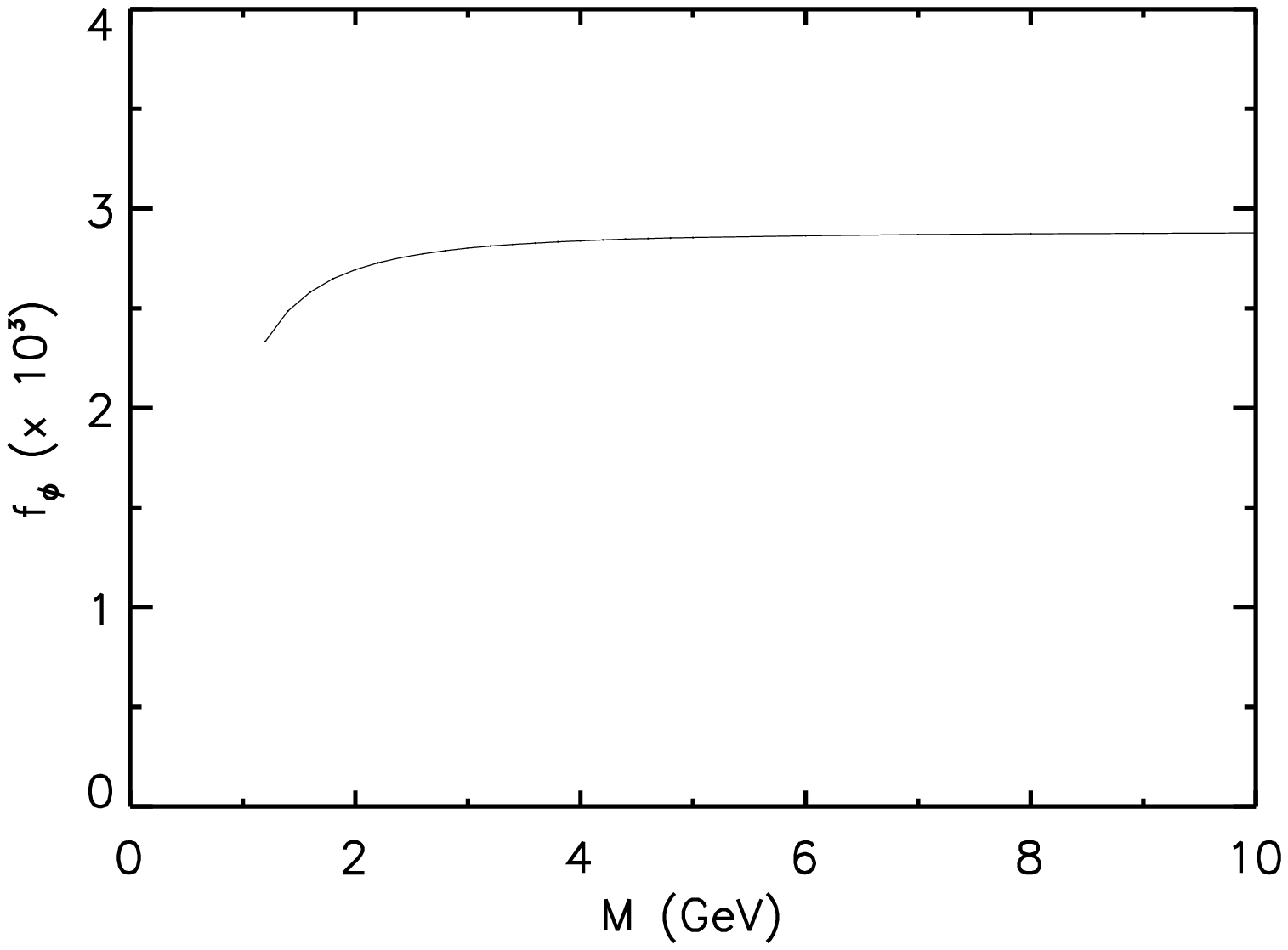,height=9cm}  }
\caption{Dependence of $f_\phi$ on the Borel mass, $M$, for modified
input set IV.}
\label{five}
\end{figure}

\begin{figure} [htb]
\centering{\
\epsfig{angle=0,figure=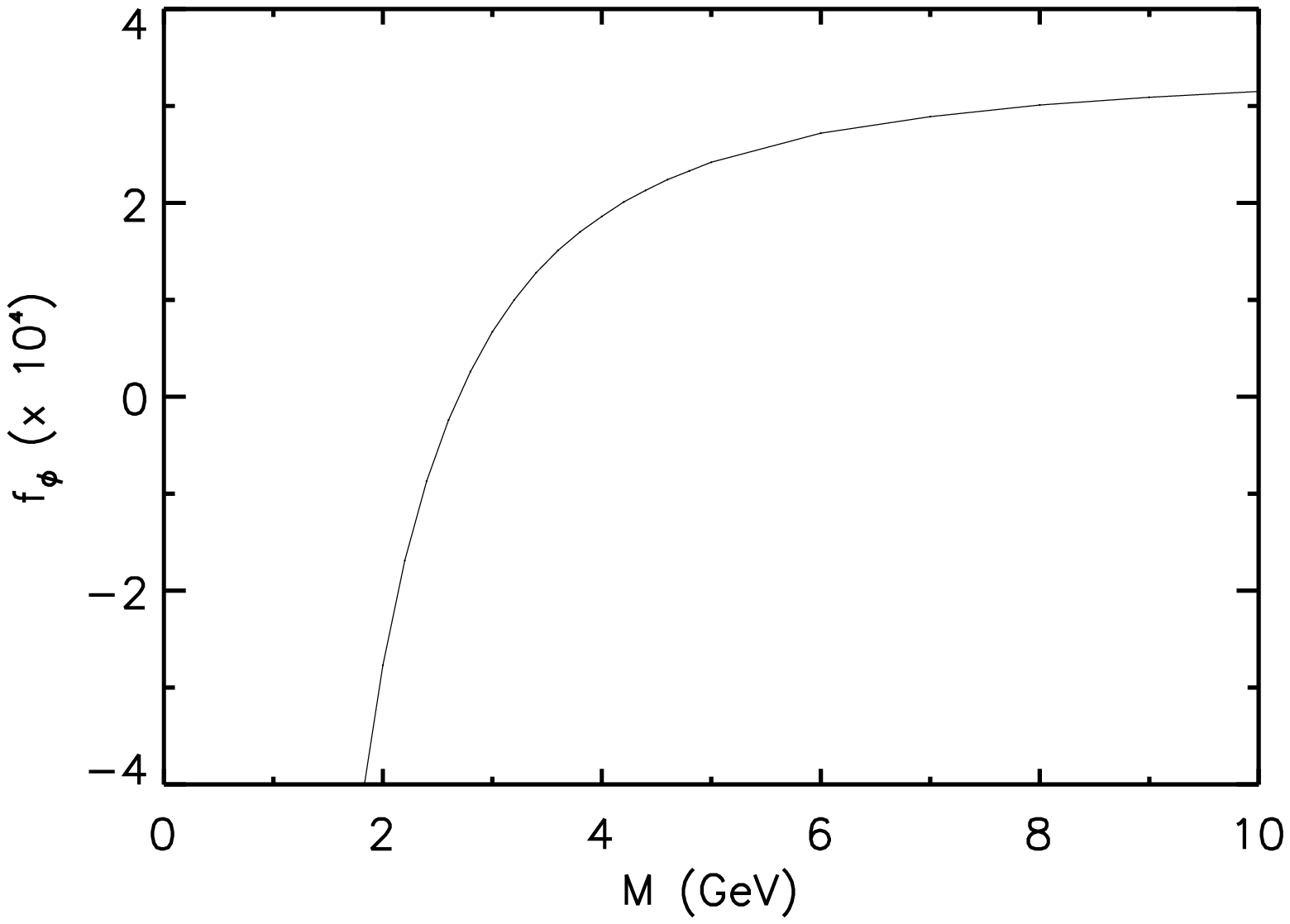,height=9cm}  }
\caption{Dependence of $f_\phi$ on the Borel mass, $M$, for modified
input set II.}
\label{six}
\end{figure}

\begin{table}
\caption{Sum rule fit for the parameters $\xi$, $\beta$, $\xi^\prime$,
$\beta^\prime$ and $f_\phi$}\label{Table1}
\begin{tabular}{lccccc}
Input&$\xi\times 10^3$&$\beta$&$\xi^\prime\times 10^5$&$\beta^\prime$&
$f_\phi\times 10^3$\\
\tableline
Set I&2.18$\pm$0.39&1.49$\pm$0.06&-2.63$\pm$0.79&-5.84$\pm$0.12&2.30$\pm$0.52\\
Set III&3.10$\pm$0.39&1.62$\pm$0.02&-4.57$\pm$0.69&-5.72$\pm$0.01&
3.57$\pm$0.52\\
Set IV&2.59$\pm$0.39&1.55$\pm$0.04&-3.47$\pm$0.61&-5.78$\pm$0.04&
2.86$\pm$0.45\\
\end{tabular}
\end{table}

\begin{table}
\caption{Sum rule fit for the isospin-breaking parameters $\{ f_k\}$.
Values are quoted for the central values of the input parameters $\{ c_i\}$.
The units are GeV$^2$.}\label{Table2}
\begin{tabular}{lccccc}
Input&$f_\rho\times 10^2$&$f_\omega\times 10^2$&$f_\phi\times 10^3$&
$f_{\rho^\prime}\times 10^4$&$f_{\omega^\prime}\times 10^4$\\
\tableline
Set I&3.53&3.73&2.30&5.34&8.45\\
Set III&5.00&5.30&3.57&9.32&14.6\\
Set IV&4.18&4.42&2.86&7.06&11.1\\
\end{tabular}
\end{table}

\begin{table}
\caption{Behavior of the correlator near $q^2=0$.  Contributions to
$12\, \Pi (0)$ from the $\rho$-$\omega$, $\phi$ and
$\rho^\prime$-$\omega^\prime$ regions are quoted for central values
of the input parameters $\{ c_i\}$ for each input set, while the
effect of the uncertainties in these values is displayed explicitly
for $12\, \Pi (0)$ and $12{d\Pi\over dq^2}(0)$.
All entries are in units of $10^{-3}$, except for $12{d\Pi\over dq^2}(0)$,
which is in units of $10^{-3}$ GeV$^{-2}$.}\label{Table3}
\begin{tabular}{lccccc}
Input&$\rho$-$\omega$&$\phi$&
$\rho^\prime$-$\omega^\prime$&
$12\Pi (0)$&$12{d\Pi\over dq^2}(0)$\\
\tableline
Set I&-1.06&2.21&-0.18&0.96$\pm$0.14&3.88$\pm$0.61\\
Set III&-1.92&3.44&-0.31&1.22$\pm$0.14&5.10$\pm$0.60\\
Set IV&-1.43&2.75&-0.24&1.08$\pm$0.14&4.43$\pm$0.61\\
\end{tabular}
\end{table}
\end{document}